\documentclass[12pt]{article}
\usepackage{epsfig}
\usepackage[numbers,sort&compress]{natbib}

\def\beq{\begin{equation}}
\def\eeq#1{\label{#1}\end{equation}}
\def\eeqn{\end{equation}}

\def\beqa{\begin{eqnarray}}
\def\eeqa#1{\label{#1}\end{eqnarray}}
\def\eeqan{\end{eqnarray}}

\let\bar=\overbar

\def\Dslash{\not{\hbox{\kern-4pt $D$}}}
\def\dslash{\not{\hbox{\kern-2pt $\del$}}}

\def\msb{{\bar{\ssstyle M \kern -1pt S}}}

\def\Title#1{\begin{center} {\Large {\bf #1} } \end{center}}

\newcommand{\vsb}{\vspace{-0.14cm}}

\def\Title#1{\begin{center} {\Large \bf #1 } \end{center}}
\def\Author#1{\begin{center}{ \sc #1} \end{center}}
\def\Address#1{\begin{center}{ \it #1} \end{center}}

\newenvironment{Abstract}{\begin{quotation}  }{\end{quotation}}
\newenvironment{Presented}{\begin{quotation} \begin{center} 
             PRESENTED AT\end{center}\bigskip 
      \begin{center}\begin{large}}{\end{large}\end{center} \end{quotation}}

\begin{document}

\begin{titlepage}

\vfill

\Title{Inclusive dihadron production at the LHC in NLA BFKL}
\vfill

\Author{F.G. Celiberto $^{a,b}$\hspace{0.05cm}\footnote{Corresponding author:  francescogiovanni.celiberto@fis.unical.it}\hspace{0.1cm}, 
D.Yu. Ivanov $^{c,d}$, 
B. Murdaca $^b$ and 
A. Papa $^{a,b}$}

\Address{
${}^a$ {\sl Dipartimento di Fisica, Universit\`a della Calabria, Arcavacata di Rende, I-87036 Cosenza, Italy}\\
${}^b$ {\sl Istituto Nazionale di Fisica Nucleare, Gruppo collegato di
Cosenza, Arcavacata di Rende, I-87036 Cosenza, Italy}\\
${}^c$ {\sl Sobolev Institute of Mathematics, RU-630090 Novosibirsk, Russia}\\
${}^d$ {\sl Novosibirsk State University, RU-630090 Novosibirsk, Russia}}
\vfill

\begin{Abstract}
A study of cross sections and azimuthal correlation moments for the inclusive production of two, light charged hadrons featuring large transverse momenta and well separated in rapidity at LHC energies, including BFKL resummation effects with NLA accuracy, is presented. This reaction shares the same theoretical setup with the well konwn Mueller-Navelet jet production and can be thought of as a new and complementary channel to access the BFKL dynamics at proton colliders.
\end{Abstract}
\vfill

\begin{Presented}
EDS Blois 2017, Prague, \\ Czech Republic, June 26-30, 2017
\end{Presented}
\vfill

\end{titlepage}
\def\thefootnote{\fnsymbol{footnote}}
\setcounter{footnote}{0}

\section{Introduction}

The investigation of semi-hard reactions in the high-energy limit is a stimulating research field in perturbative QCD, the Large Hadron Collider (LHC) yielding us a wealth of useful data. In this kinematical regime, the Balitsky-Fadin-Kuraev-Lipatov (BFKL) resummation~\cite{Fadin:1975cb,Kuraev:1976ge,Kuraev:1977fs,Balitsky:1978ic} represents the most effective mechanism to account for the large logarithms in the colliding energy which are present to all orders of the perturbative expansion, with leading (LLA) and next-to-leading (NLA) accuracy. The inclusive production of two jets with high $p_T$ and well separated in rapidity, namely the Mueller-Navelet jet reaction~\cite{Mueller:1986ey}, 
has been one of the most studied processes in the last years. In this case, the BFKL resummation in the NLA reckons on the convolution of the NLA Green's function of the BFKL equation~\cite{Fadin:1998py,Ciafaloni:1998gs} with the NLA jet impact factors~\cite{Bartels:2001ge,Bartels:2002yj,Caporale:2011cc,
               Ivanov:2012ms,Colferai:2015zfa}.
In~\cite{Colferai:2010wu,Angioni:2011wj,Caporale:2012ih,
         Ducloue:2013hia,Ducloue:2013bva,Caporale:2013uva,
         Ducloue:2014koa,Caporale:2014gpa,Ducloue:2015jba,
         Celiberto:2015yba,Celiberto:2016ygs,Chachamis:2015crx,N.Cartiglia:2015gve}  
NLA BFKL predictions for cross sections and azimuthal-angle observables~\cite{DelDuca:1993mn,Stirling:1994he,Vera:2006un,Vera:2007kn} for the Mueller-Navelet jet process were presented, showing a very nice agreement with experiment~\cite{Khachatryan:2016udy}.
Pursuing the goal to give a deeper and more complete description of the dynamics behind strong interactions in the Regge limit, $s\gg |t|$, some other, BFKL-sensitive observables have to be considered in the context of the LHC physics plans. 
A fascinating option, the study of three- and four-jet production processes, separated by a large rapidity interval from each other, was recently proposed in~\cite{Caporale:2015vya,Caporale:2015int} and examined  in~\cite{Caporale:2016soq,Caporale:2016zkc,Caporale:2016xku,Celiberto:2016vhn}.
A novel possibility, {\it i.e.} 
the inclusive production of two, light charged  hadrons: $\pi^{\pm}$, $K^{\pm}$, $p$, $\bar p$ 
featuring high $p_T$ and well separated in rapidity, together with an undetected gluon radiation emission is investigated in this paper. 
As well as for Mueller-Navelet jets, NLA BFKL treatment is applicable to this process, since the impact factor for the production of an identified hadron was calculated with NLA accuracy in~\cite{hadrons}. 
On one hand, LHC detectors allow us to tag hadrons at much smaller $p_T$ values than jets, opening the doors of a kinematic region complementary to the one studied with Mueller-Navelet analyses. 
On the other hand, we can use this reaction as a powerful tool to constrain not only the parton distributions (PDFs) for the initial proton, but also the parton fragmentation functions (FFs) 
describing the tagged hadron in the final state.

\section{Inclusive dihadron production}

\begin{figure}[t]
\centering

   \includegraphics[scale=0.34,clip]{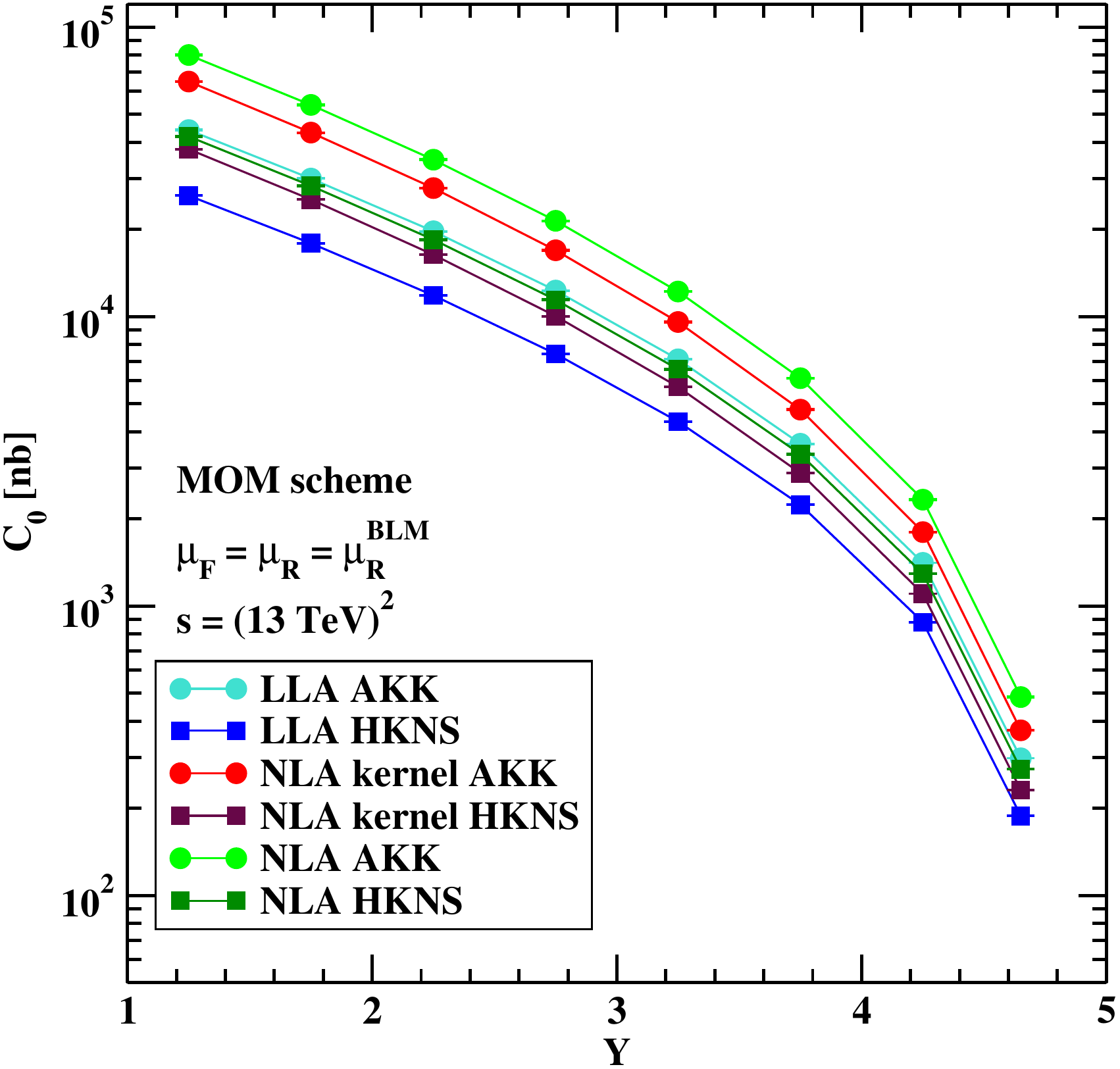}
   \includegraphics[scale=0.34,clip]{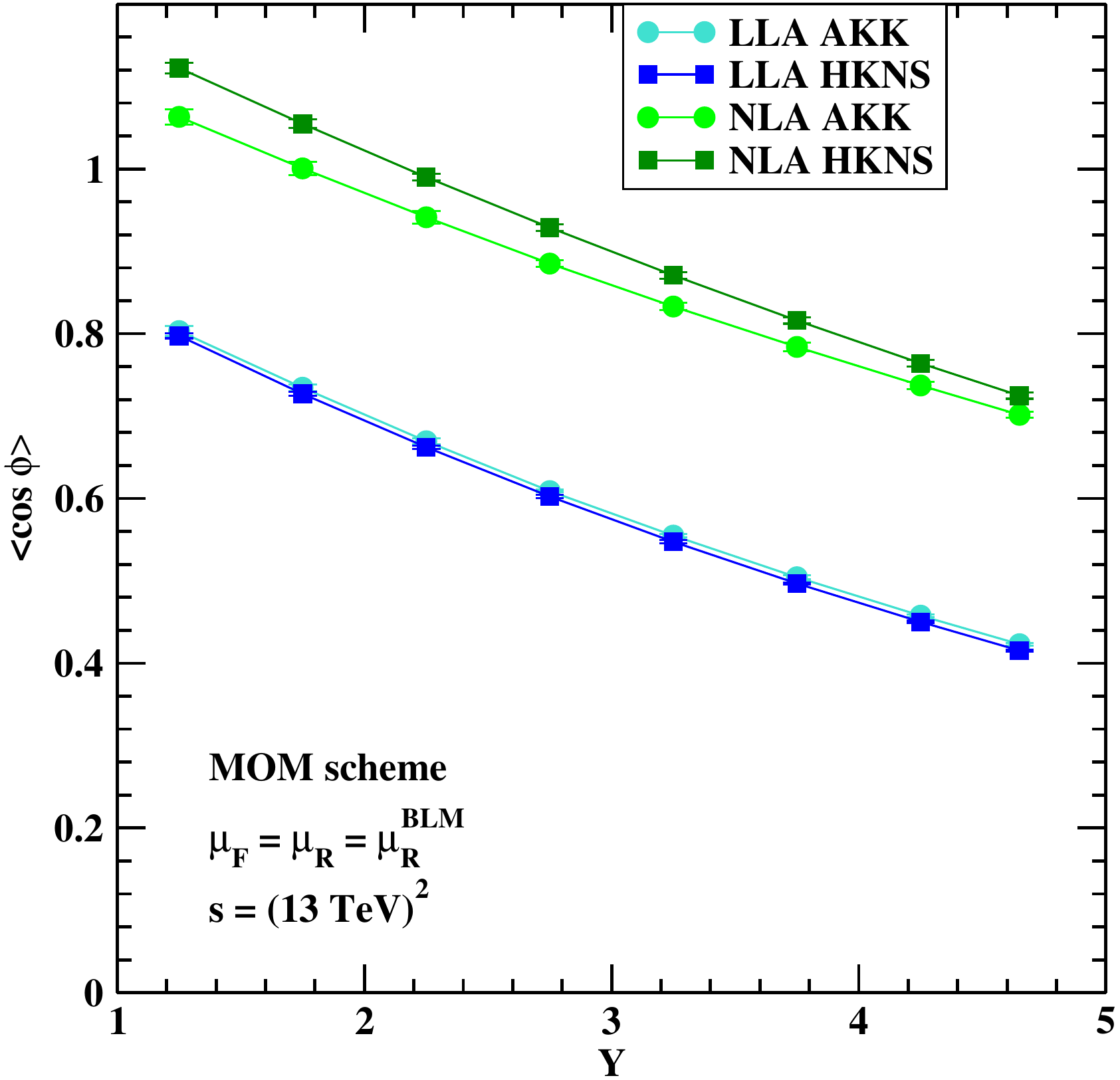}

   \includegraphics[scale=0.34,clip]{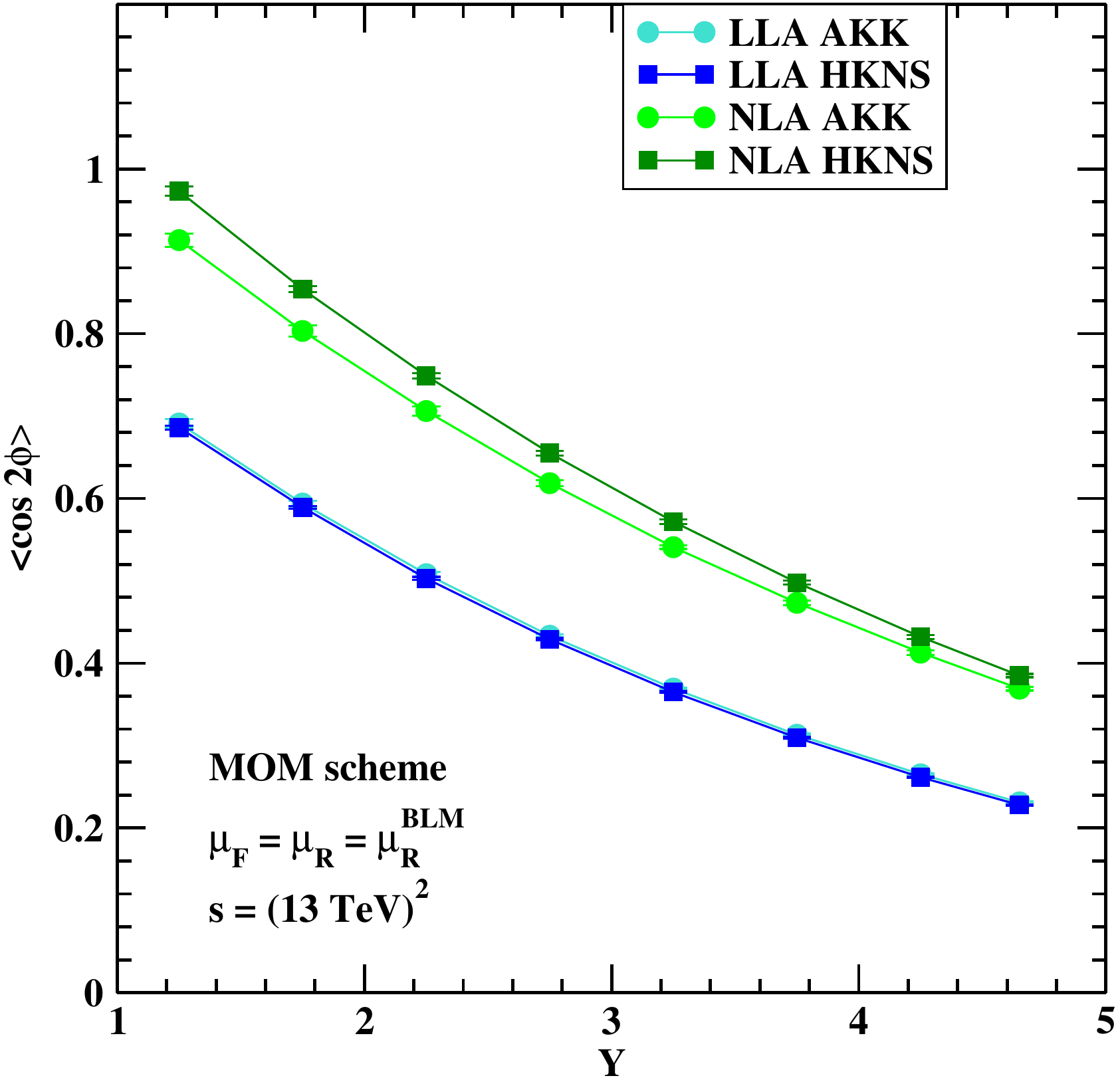}
   \includegraphics[scale=0.34,clip]{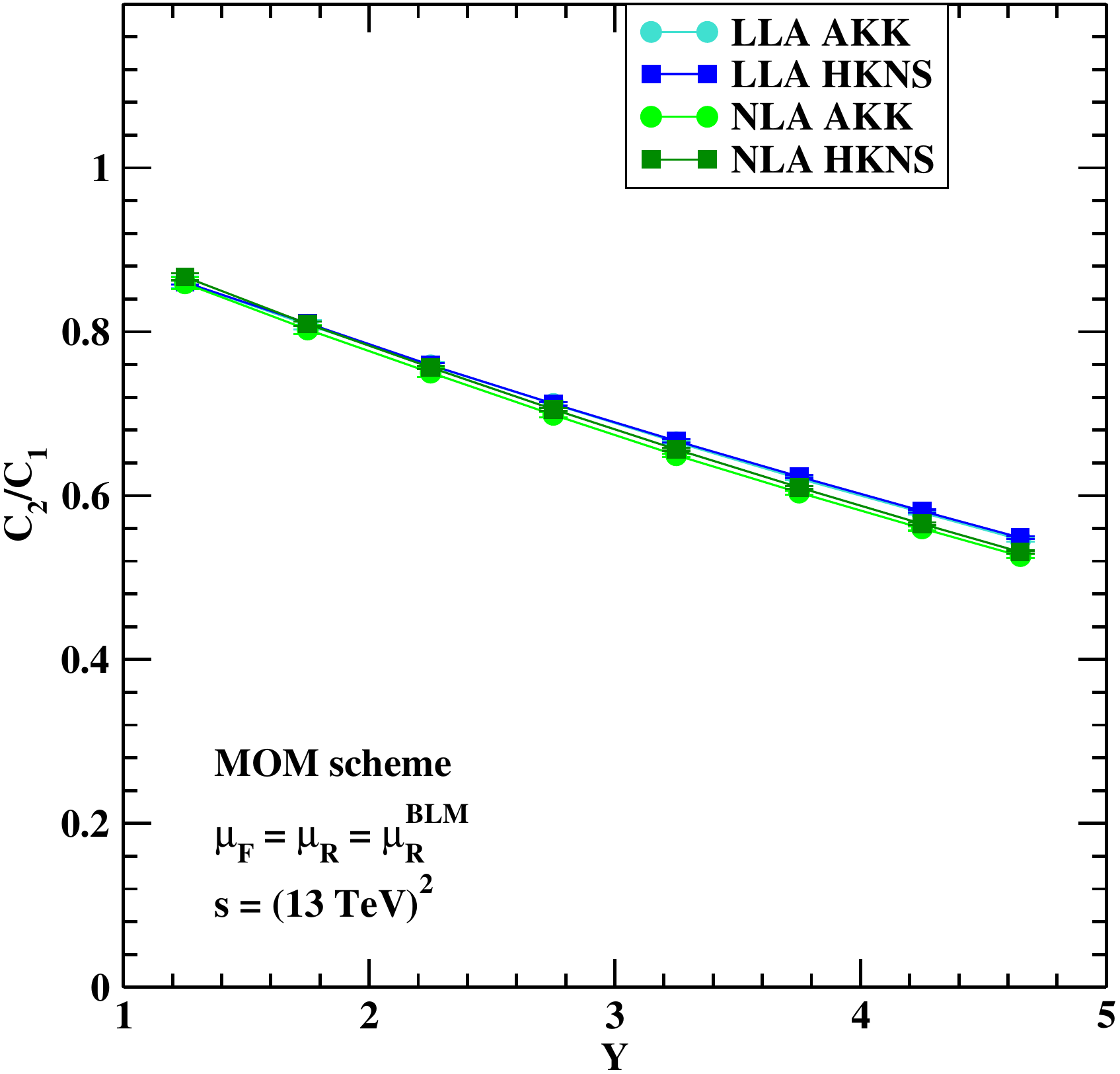}

\caption{$Y$-dependence of $C_0$ and of several ratios $R_{nm}$ at $\sqrt{s} = 13$ TeV.}
\label{fig:plots}
\end{figure}

We study the inclusive hadroproduction
of two identified hadrons in proton-proton collisions
\begin{eqnarray}
\label{eq:process}
{\rm p}(p_1) + {\rm p}(p_2) \to {\rm h}(k_1, y_1, \phi_1) 
                              + {\rm h}(k_2, y_2, \phi_2) 
                              + {\rm X} \;,
\end{eqnarray}
where the hadrons have high transverse momenta, 
$\vec k_1^2\sim \vec k_2^2 \gg \Lambda^2_{\rm QCD}$ and large separation in rapidity $Y=y_1-y_2$. The differential cross section of the process reads
\begin{equation}
\label{eq:dcs}
\frac{d\sigma}{dy_1dy_2\, d|\vec k_1| \, d|\vec k_2|d\phi_1 d\phi_2}
=
\frac{1}{(2\pi)^2}
\left[
{\cal C}_0+\sum_{n=1}^\infty  2\cos (n\phi ) \, {\cal C}_n \right] \;,
\end{equation}
where $\phi=\phi_1-\phi_2-\pi$, with $\phi_{1,2}$ the two hadrons'
azimuthal angles, while ${\cal C}_0$ gives the total
cross section and the other coefficients ${\cal C}_n$ determine 
the azimuthal angle distribution of the two hadrons. 
In order to match the kinematic cuts 
used by the CMS collaboration, we consider 
the \emph{integrated coefficients}
given by
\begin{equation}
\label{eq:Cm_int}
C_n=
\int_{y_{1,\rm min}}^{y_{1,\rm max}}dy_1
\int_{y_{2,\rm min}}^{y_{2,\rm max}}dy_2
\int_{k_{1,\rm min}}^{\infty}dk_1
\int_{k_{2,\rm min}}^{\infty}dk_2
\; \delta\left(y_1-y_2-Y\right)
\; {\cal C}_n
\end{equation}
and their ratios $R_{nm}\equiv C_n/C_m$.
As for the integration range, we consider:
$y_{1,2, \rm max}=-y_{2,1, \rm min}=2.4$, with 
and $Y \leq 4.8$, and $k_{1,2, \rm min}=5$~GeV. We use the MSTW 2008 NLO~\cite{Martin:2009iq} 
PDF set and two different NLO parameterizations for hadron FFs:  
AKK~\cite{Albino:2008fy} and HKNS~\cite{Hirai:2007cx}.
As for the renormalization and the factorization scales\footnote{For more detailed studies on the use of different choices for the values of the scales, see~\cite{Celiberto:2016hae,Celiberto:2017ptm,Celiberto:2017ius}.}, we take $\mu_F=\mu_R$ and use the Brodsky-Lepage-Mackenzie (BLM) 
scheme~\cite{Brodsky:1982gc} as derived in its ``exact'' version in~\cite{Caporale:2015uva}. All calculations are done in the MOM renormalization scheme.

Results for $C_0$ and for several ratios $R_{nm}$ at $\sqrt{s}=13$ TeV are shown in Figure~\ref{fig:plots}.

\section{Conclusions \& Outlook}

We studied the dihadron production
at the LHC, presenting the first theoretical predictions for cross sections and azimuthal angle correlations in the NLA BFKL accuracy. 
The general features of our predictions for dihadron production are rather similar to those obtained earlier for the Mueller-Navelet jet process. In particular, we observe that the account of NLA BFKL terms leads to much
less azimuthal angle decorrelation with increasing $Y$ in comparison to LLA BFKL calculations. 

We plan to extend this study by considering more exclusive final-state processes. If, together with the hadron, a forward jet is also emitted, we will
have the chance to investigate \emph{hadron-jet} correlations. The study of another interesting reaction, the inclusive production of two heavy quark-antiquark pairs, well separated in rapidity, in ultra-peripheral collisions at the LHC is in progress~\cite{Celiberto:2017nyx}.

\end{document}